\documentclass[prd,11pt]{article}

\usepackage{dsfont}
\usepackage{geometry}
\usepackage{authblk}
\usepackage{latexsym}

\usepackage{enumerate}
\usepackage[inline]{enumitem}
\usepackage{amsmath,amssymb}
\usepackage{amsfonts,dsfont}
\usepackage{nicefrac}
\usepackage{microtype}
\usepackage{mathtools}

\usepackage{sidecap}
\usepackage{caption}
\usepackage{subcaption}
\usepackage{wrapfig}
\usepackage{enumitem}
\usepackage{algorithm}
\usepackage[normalem]{ulem}
\usepackage{amssymb}
\usepackage{multicol}
\usepackage{adjustbox}
\usepackage{multirow}
\usepackage{color}
\usepackage{xspace}
\usepackage{CJKutf8}

\PassOptionsToPackage{numbers}{natbib}
\usepackage{natbib}

\usepackage[colorinlistoftodo   s,prependcaption,textsize=tiny]{todonotes}

\usepackage{user_definitions}
\usepackage{xspace}
\usepackage{algorithm}
\usepackage{algorithmicx}
\usepackage[no end]{algpseudocode} 
\usepackage{amsmath,amssymb}
\usepackage{mathtools}
\usepackage{sidecap}
\usepackage{booktabs}
\algdef{SE}[DOWHILE]{Do}{doWhile}{\algorithmicdo}[1]{\algorithmicwhile\ #1}%

\newcommand{\dataset}{X}

\newcommand{\tree}{\mathtt{H}}
\newcommand{\treeset}{\mathcal{H}}

\newcommand{\EfunS}{\ensuremath{\psi}}
\newcommand{\EfunT}{\ensuremath{\phi}}

\def\beq{\begin{equation}}
\def\eeq{\end{equation}}
\newcommand{\bea}{\begin{eqnarray}\begin{aligned}}
\newcommand{\eea}{\end{aligned}\end{eqnarray}}

\usepackage[american]{babel}
\usepackage{natbib} % has a nice set of citation styles and commands
\usepackage{mathtools} % amsmath with fixes and additions
\usepackage{booktabs} % commands to create good-looking tables
\usepackage{tikz} % nice language for creating drawings and diagrams

\newcount\Comments  % 0 suppresses notes to selves in text
\definecolor{darkgreen}{rgb}{0,0.5,0}
\definecolor{darkred}{rgb}{0.7,0,0}
\definecolor{teal}{rgb}{0.1,0.6,0.7}
\definecolor{blue}{rgb}{0.0,0.1,0.9}
\definecolor{orange}{rgb}{1.,0.7,0.0}
\definecolor{lightblue}{rgb}{0.70, 0.80, 0.89}
\newcommand{\kibitz}[2]{\ifnum\Comments=1{{\textcolor{#1}{\textsf{\footnotesize [#2]}}}}\fi}

\usepackage{hyperref}       % hyperlinks
\usepackage{url}            % simple URL typesetting

\title{The Quantum Trellis: A classical algorithm for sampling the parton shower with interference effects}

\begin{document}
\author[]{Sebastian Macaluso}
\author[]{Kyle Cranmer}

\date{} 
\affil[]{\small{Center for Cosmology and Particle Physics \& Center for Data Science,\\ New York University, USA}}

\maketitle
\begin{abstract}
% \vspace{-.2cm}

Simulations of high-energy particle collisions, such as those used at the Large Hadron Collider, are based on quantum field theory; however, many approximations are made in practice. For example, the simulation of the parton shower, which gives rise to objects called `jets', is based on a semi-classical approximation that neglects various interference effects. While there is a desire to incorporate interference effects, new computational techniques are needed to cope with the exponential growth in complexity associated to quantum processes. We present a classical algorithm called the \textit{quantum trellis} to efficiently compute the un-normalized probability density over $N$-body phase space including all interference effects, and we pair this with an MCMC-based sampling strategy. This provides a potential path forward for classical computers and a strong baseline for approaches based on quantum computing. 

\end{abstract}

% \vspace{-.5cm}
\section{Introduction}
% \vspace{-.1cm}

The high-energy particle physics community relies on high-fidelity predictions 
of particle collisions. These predictions are based on quantum field theory, but in practice many approximations are made. While tools like \texttt{MadGraph}~\citep{Alwall:2011uj} model the \textit{hard} collision and include interference effects between different Feynman diagrams, the parton shower implemented in tools like \texttt{Pythia}~\citep{Sjostrand:2006za} are based on a semi-classical approximation that neglects various interference effects. In particular, the classical treatment of the parton shower admits an efficient sampling algorithm where the shower evolves sequentially and is described by an autoregressive probabilistic model (a Markov process). There is a desire to improve upon the classical treatment and explicitly incorporate interference effects~\citep{Nagy:2008eq,Nagy:2014mqa}; however, new computational techniques are needed to cope with the exponential growth in complexity associated to quantum processes~\citep{Provasoli:2019tzz, Bauer:2019qxa,Bauer:2021gup}.

%\newpage 
\textbf{Contributions of this paper} We present a classical data structure (the \textit{quantum trellis}) and dynamic programming algorithm to efficiently compute the un-normalized probability density over $N$-body phase space including all interference effects and pair this with a MCMC-based sampling strategy. This provides a potential path forward for sampling the parton shower including interference effects with classical computers and a strong baseline for approaches based on quantum computing.

\section{The classical trellis}

\begin{wrapfigure}{r}{0.4\textwidth} 
\vspace{-20pt}
  \begin{center}
    \includegraphics[width=0.25\textwidth]{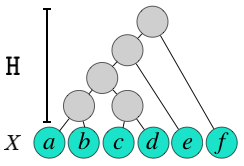}
    % \vspace{-0.75cm}
    \caption{\small{Schematic representation of a hierarchical clustering $\tree$ for the dataset $X$. }}
    \label{fig:latentStructure}
  \end{center}
  \vspace{-0.6cm}
%   \vspace{1pt}
\end{wrapfigure}

The hierarchical trellis described in~\cite{pmlr-v130-macaluso21a} is a data structure that can be paired with a dynamic programming algorithm to efficiently search or sum over the enormous space $\mathcal{H}$ of hierarchical clusterings  of $N$ objects. It generalizes a previously developed algorithm described in \cite{NIPS2018_8081} for flat clustering. The hierarchical generalization was motivated by the study of `jets' at the LHC~\citep{Cranmer:2021gdt}. In that context, the $N$ objects to be clustered correspond to final state particles observed in the large particle detectors like ATLAS and CMS. Each of those particles have energy and momentum as features $x_i \in \mathbb{R}^4$. The hierarchical clusterings $\tree \in \mathcal{H}$ correspond to one of the possible latent showering histories that could give rise to $X=\{x_i\}_{i=1}^N$ via the parton shower (see Figure \ref{fig:latentStructure}). Many tasks in jet physics involve either searching for the most likely showering history $\hat{\tree}(X)$ or calculating the marginal likelihood by summing over all possible showering histories. The computational difficulty lies in the fact that there are $(2N-3)!!$ possible hierarchical clusterings and an additional $2^{N-1}$ permutations of the left/right children at each binary splitting\footnote{See Refs.~\cite{callan2009combinatorial,DaleMoonCatalanSets} for more details and proof. }. Therefore, brute-force search or sum is not feasible for even $N \approx 10$.

Ref.~\cite{pmlr-v130-macaluso21a} considered a so-called ``energy-based'' probabilistic model for hierarchical clustering, though ``energy'' is not to be taken literally in the physics context\footnote{This is referred to as an energy-based model since often it is the case that $\EfunS(\cdot, \cdot)$ has the form of an unnormalized Gibbs or Boltzmann distribution, as $\EfunS(z_L, z_R) = \exp(-\beta E(z_L,z_R))$, where $\beta$ is the inverse temperature and $E(\cdot,\cdot)$ is the energy. }. 
In particular, the hierarchical trellis and dynamic programming algorithms were specialized to situations where the unnormalized probability $\EfunT(\dataset | \tree)$ is given by a product over terms for each of the splittings, where $\EfunS(z_L, z_R)$ is proportional to the probability the parent split into the corresponding left- and right- sibling nodes.
The posterior probability $P(\tree|\dataset)$ of $\tree$ for the dataset
   $\dataset$ is equal to the unnormalized potential
  of $\tree$ normalized by the partition function,
  $Z(\dataset)$:
   \vspace{0.55cm}
  \vspace{-.2cm}
  \begin{equation}\label{eq:potential}
      P(\tree|\dataset) = \frac{\EfunT(\dataset | \tree)}{Z({\dataset})} \,\,\,\,\,\,  \text{ with } \,\,\,\,\,\, \EfunT(\dataset | \tree) = \prod_{z_L,z_R \in \textsf{siblings}(\tree)} \EfunS(z_L,z_R)
  \end{equation} 
where $z_{L/R} = x_{L/R}$ for the leaves, the partition function $Z(\dataset)$ is given by
\vspace{-.1cm}
\begin{equation}
    Z({\dataset}) = \sum_{\tree \in \treeset({\dataset})} {\EfunT(\dataset | \tree)} \label{eq:Z}
\end{equation}
% \vspace{-.1cm}
and $\treeset({\dataset})$ gives all binary hierarchical clusterings of the elements $\dataset$.

The resulting algorithm's computational complexity is $\mathcal{O}(3^{N})$. While still exponential, it is a super-exponential improvement over a naive iteration over the hierarchies $\tree \in \mathcal{H}$. 
This makes it feasible in regimes where enumerating all possible trees would be infeasible, and is to our knowledge the fastest exact partition function result, making practical exact inference for datasets on the order of 20 elements (%\textasciitilde 
$\approx 3\times10^9$ operations vs. %\textasciitilde 
$\approx 10^{22}$ trees).

\section{The shower model}

\paragraph{Classical Ginkgo}
The autoregressive form of the potential function 
$\EfunT(\dataset | \tree)$ in Eq.~\ref{eq:potential} as a product over splittings where each splitting contributes $\EfunS(z_L, z_R)$ is consistent with the classical description of the parton shower. 
However, in practice state-of-the-art parton showers do not expose the splitting likelihoods in a convenient way. 
Thus, to aid in machine learning research for jet physics, a python package for a simplified generative model of a parton shower, called \texttt{Ginkgo}, was introduced in \cite{ToyJetsShowerPackage, ginkgo}.
\texttt{Ginkgo} exposes  the probability model for each splitting and has a tractable joint likelihood $p(X | \tree)$. 
Each edge in the tree corresponds to a particle with an energy-momentum vector $z = (E \in \RR^{+}, \vec{p} \in \RR^3)$ and squared mass ${t}(z) = {E^2-|\vec{p}|^2}$. 
The energy-momentum vector is conserved in a splitting, i.e., $z_P = z_L + z_R$. The decay of the parent into children is isotropic in the parent's rest frame and the squared mass $t(z)$ of the children is distributed according to a truncated exponential distribution 
    \begin{equation}\label{eq:ginkgo_split_p}
        f(t | t_{\rm max}, \lambda) = \frac{1}{1 - e^{- \lambda}} \frac{\lambda}{t_{\rm max}} e^{- \lambda \frac{t}{t_{\rm max}}} \;,
    \end{equation}
    %\end{align}
where the first term is a normalization factor associated to the constraint that $t<t_{\rm max}$.
Care must be taken to avoid generating configurations where the mass of the children is greater than the mass of the parent. This is done by sequentially generating the two children, i.e. $t_L \sim f(t_L|t_P,\lambda)$ and then  $t_R \sim f(t_R |  t_P^{LR}, \lambda)$ (or vice versa with $t_R \sim f(t_R | t_P, \lambda)$ and then $t_L \sim f(t_L | t_P^{RL}, \lambda)$), where $t_P^{LR/RL}= (\sqrt{t_P} - \sqrt{t_{L/R}})^2$ imposes necessary boundary conditions. The final model is then a symmetric mixture of the two orderings: 
\begin{equation}
\EfunS(z_L, z_R) = \frac{1}{2}[{f(t_L | t_P, \lambda)} { f(t_R | t_P^{LR}, \lambda)}+{f(t_R | t_P, \lambda)} { f(t_L | t_P^{RL}, \lambda)}] \;.
\end{equation}
When the shower terminates the latent variables $\{ z_i\}_{i \in \textrm{leaves}}$ are identified with the set of observed random variable $X$. The final distribution $p(X)$ is invariant to permutations. 

While \texttt{Ginkgo} is a simplification of modern parton showers generators, it captures essential ingredients of the physical process. 
Within the analogy between jets and natural language processing (NLP)~\cite{Louppe:2017ipp,andreassen2019junipr}, \texttt{Ginkgo} can be thought of as a generative language model that produces $(X,\tree)$ pairs where $X$ is the text, $H$ is the ground-truth parse trees, and $p(X|\tree)$ is known.

\paragraph{The analogous quantum mechanical amplitude}
In order to study the quantum analogue of the \texttt{Ginkgo} parton shower model, we must have a quantum mechanical amplitude $\mathcal{A}(X|\tree)$ for a given hierarchy $\tree$. 
We base our amplitude on the classical generative model implemented in \texttt{Ginkgo}.
In keeping with what one expects from Feynman rules, we define the amplitude of a hierarchy $\mathcal{A}(X|\tree)$ as the product of the amplitudes $\mathcal{A}(z_L, z_R)$ for all the  $1\rightarrow 2$ splittings, which includes a complex phase that depends on the invariant mass of the parent. Specifically, we choose
    \begin{align}\label{eq:quantum_ginkgo}
        \mathcal{A}(z_L, z_R | \beta)= %\frac{1}{2} 
        e^{-i \beta t_P}
        \sqrt{\EfunS(z_L, z_R)}
    \end{align}
    where the first term is a $t_P$-dependent complex phase with hyperparameter $\beta$ and the square-roots are introduced to maintain consistency with the splitting likelihoods used in the classical Ginkgo model in Eq.\ref{eq:ginkgo_split_p}.
Note that in the case where $\beta=0$ each term is real, but there is still constructive interference when calculating $|\mathcal{A}(X|\tree)|^2$. Therefore $\beta=0$ does not correspond to the classical case. We would like to end this section by emphasizing that the quantum mechanical amplitude defined by Eq.~\ref{eq:quantum_ginkgo} is not meant to be physically justified, but to have the right form for exploring efficient algorithms for parton showers with quantum interference.

\section{Quantum Hierarchical trellis}\label{quantum_trellis}

The basic idea for the \textit{quantum trellis} is simple: replace the real-valued potential function $\EfunS(z_L, z_R)$ with the complex-valued amplitude $\mathcal{A}(z_L, z_R)$. Each path through the trellis will correspond to a particular hierarchy $\tree$, and one can accumulate the terms to compute $\mathcal{A}(X|\tree) = \prod_{z_L, z_R \in \textrm{siblings}(\tree)} \mathcal{A}(z_L, z_R)$ as before. Then we can use the same hierarchical trellis data structure and dynamic programming algorithm to efficiently compute the total amplitude for all possible $(2N-3)!!$ showering histories $\mathcal{A}(X) = \sum_{\tree \in \mathcal{H}(X)} \mathcal{A}(X|\tree)$.\footnote{The Cluster Trellis code can be accessed at \href{https://github.com/SebastianMacaluso/ClusterTrellis}{https://github.com/SebastianMacaluso/ClusterTrellis} } 

\begin{figure}[h]
        \centering  
        \includegraphics[width=0.95\textwidth]{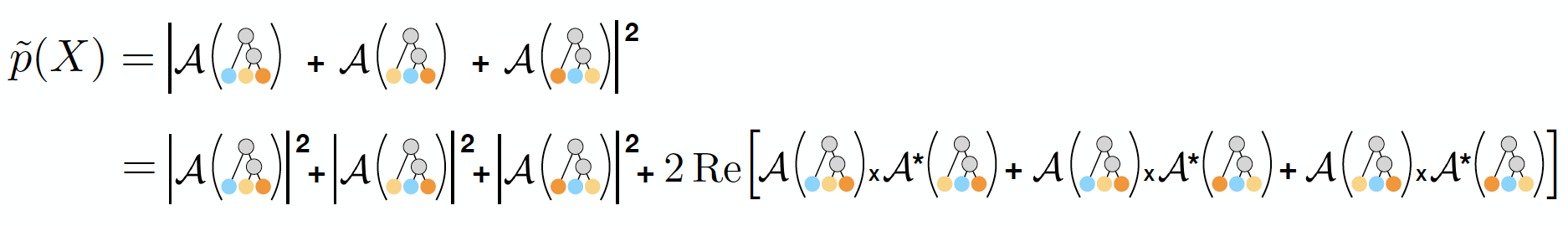} 
    \vskip -0.1in
    \caption{\small{Schematic representation of all the terms of the marginal likelihood for a dataset of three elements, showing the interference terms.}}
    \vspace{-0.4cm}
\label{fig:trellis_example_b}
\end{figure}

Using the Born rule, the un-normalized probability density $\tilde{p}(X)$ is given by the square of the magnitude of the total amplitude

\begin{eqnarray}
\tilde{p}(\dataset) = \bigg| \sum_{\tree \in \treeset({\dataset})} {\mathcal{A}(\dataset | \tree)}  \bigg|^2 \, p(z_{\rm root})\; ,%= | Z(\dataset) |^2
\end{eqnarray}
where $p(z_{\rm root})$ is a distribution of latent variables associated to the root node. In \texttt{Ginkgo} we sample the root node 3-momentum and invariant mass of the root from a multivariate normal distribution. Note due to energy and momentum conservation, $z_{\rm root}$ is equal to the sum over all the leaves and is independent of $\tree$. The distribution $\tilde{p}(\dataset)$ includes all constructive and destructive interference among the different hierarchies. In principle, this includes all $((2N-3)!!)^2$ cross-terms as schematically illustrated in Figure~\ref{fig:trellis_example_b}; however, one need not explicitly construct cross terms $\mathcal{A}_i \mathcal{A}^*_j$ if one simply performs the sum before squaring. 

 To the best of our knowledge, this is the first time the marginal amplitude can be exactly obtained over datasets of $\mathcal{O}(10)$ elements. For example, with $N=8$ there are 135,135 possible hierarchies and over 18 billion cross terms.

\section{Sampling including interference effects}

The final goal is to be able to generate events according to the distribution including all interference effects, e.g. we want to sample $X \sim {p}(X)$. In typical parton showers like \texttt{Pythia} and the classical \texttt{Ginkgo}, which use a semi-classical approximation, the generative model proceeds as a Markov process starting at the root node, sampling left and right children $z_L, z_R \sim p(z_L, z_R | z_P)$ recursively. At each step of the recursion, there is also a probability that the parent node will not split and become a leaf node. Sampling is fairly easy at each step since the distribution at each splitting is not very high dimensional. In the end, the latent variables associated to internal nodes are ignored and the data associated to the leaves serve as samples for the implicit marginal distribution. In the quantum case this strategy won't work because the interference effects destroy the conditional independence of the individual splittings. 

Instead, we seek a sampling strategy that operates directly on the phase space of the observed particles (the leaves) instead of the internal latent variables.
We employ Markov Chain Monte Carlo (MCMC) techniques, which only need the target distribution $\tilde{p}(X)$ defined up to a multiplicative constant. 
We use the  \href{https://emcee.readthedocs.io/en/stable/}{emcee} library from \cite{Foreman_Mackey_2013} and fix the number of final state particles $N$, which corresponds to $X\in \mathbb{R}^{4N}$ phase space. 

The typical (semi-classical) parton shower provides not only distribution over phase space, but also the distribution over the number of final state particles (leaves) $N$. 
Correctly generating the distribution over $N$ in the quantum setting is a challenging problem as it involves the phase space integrals $\int dX\, \tilde{p}(X|N)$ and is left for future work.

\section{Experiments: Jet Physics}\label{experiments}

\begin{figure}
% \centering
\raggedleft
    \begin{minipage}{0.45\textwidth}
        \centering
        \includegraphics[width=\textwidth]{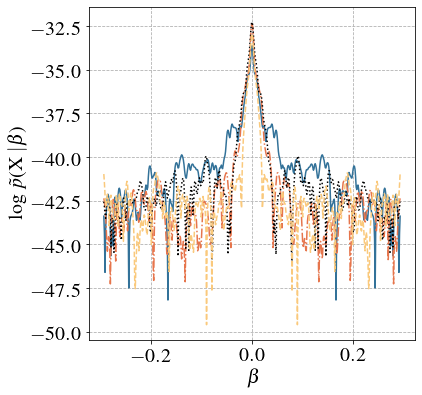}
        \vspace{-.3cm}
        \caption{\small{Log of the un-normalized likelihood $\tilde{p}(X|\beta)$ vs the hyperparameter $\beta$ for four independent $X$ with 8 leaves sampled from the classical Ginkgo model. We see that all datasets have a similar behavior, with a peak at zero where there is only constructive interference.  
        }}
        \label{fig:QuantumTrellis_beta}
    \end{minipage}\hfill
    \begin{minipage}{0.45\textwidth}
  \centering
    \includegraphics[width =\textwidth]{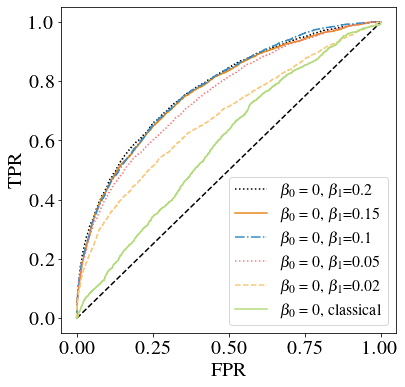}
    \vspace{-.5cm}
      {\caption{\small{ROC curves between pairs of datasets (with 8 leaves) sampled with MCMC for different values of $\beta$. We also show the ROC curve between the $\beta=0$ model and classical Ginkgo (solid green).
}}
\label{fig:ROC}}
    \end{minipage}
    \vspace{-.5cm}
\end{figure}

In  Figure~\ref{fig:QuantumTrellis_beta}, we show the dependence of $\tilde{p}(X|\beta)$ on the hyperparameter $\beta$ (see Eq. \ref{eq:quantum_ginkgo}) for four independent $\dataset$ sampled from the classical \texttt{Ginkgo} model \footnote{The \texttt{Ginkgo} parameters for these samples are $\lambda=1.5$, the mass of the root node was sampled from  a normal distribution with mean 30~GeV and standard deviation 5 $\text{GeV}$, and each component of the root node's 3-momentum vector was sampled with a normal distribution with mean 200 GeV and standard deviation 10 GeV.}. We see that each has a similar behavior with the un-normalized likelihood peaking at $\beta=0$ where we only have  constructive interference.%, as expected.

Using the MCMC technique described above we generated samples for different values of the hyperparameter $\beta$. For each $\beta$ we generated a set of 260000 samples.
These sets were obtained from running MCMC for 3250 steps and 80 walkers (with an additional 100 burn-in steps).
It took about 78 hours to generate each set of 260000 samples on an Intel Xeon Platinum 8268 24C 2.9GHz Processor.

Next, we characterize the effect of interference via the ROC AUC between datasets $X\sim p(X | \beta=0)$ and $X\sim p(X | \beta=\beta_1)$, for different values of $\beta_1$.  
With the trellis algorithm, we are able to directly calculate the discrimination power of the optimal classifier without training a classifier
\footnote{ The optimal classifier is based on the Neyman–Pearson lemma and defined by the likelihood ratio as the most powerful variable or test statistic (for a proof and a particle physics application see \cite{optimalClassifier,Cranmer_2007}).}. 
We show in Figure~\ref{fig:ROC} the ROC curve for $\tilde{p}(X)$, between pairs of datasets with 8 leaves for different values of $\beta_1$.
We also show the ROC curve between the $\beta=0$ model and classical \texttt{Ginkgo}. 
We can see that MCMC together with the quantum trellis allows to generate samples that are different from the ones generated with classical \texttt{Ginkgo}. 

\section{Conclusion}
% \vspace{-0.25cm}

We developed the quantum trellis data structure and dynamic programming  algorithm that allows us to efficiently calculate the probability using the Born rule, as well as a MCMC technique to sample from $\tilde{p}(X)$ including all interference effects. The resulting approach provides a strong classical baseline for potential quantum algorithms that might be used to sample from a parton shower including interference effects.

We end by noting that \cite{Provasoli:2019tzz, Bauer:2019qxa, Bauer:2021gup}) considered quantum algorithms for simulating a similar system: the binary random walk of a single particle taking $N$ steps to the left or right. In that case, the state space looks like a binary tree with $2^N$ leaves, but each realization is a single path. They developed efficient algorithms to sample the quantum  process where multiple paths interfere. Our case is exponentially harder as each realization is not a single path but itself a binary tree, thus the state space corresponds to the $(2N-3)!!$ showering histories for $N$ particles.

\section*{Acknowledgements}

Kyle Cranmer and Sebastian Macaluso are supported by the National Science Foundation under the awards ACI-1450310 and OAC-1836650 and by the Moore-Sloan data science environment at NYU. 

\bibliography{references}
\end{document}